\documentclass[onecolumn,secnumarabic,amssymb, amsmath, nofootinbib,tightenlines,
nobibnotes, aps, prl,epsfig]{revtex4}
\usepackage{graphicx}% Include figure files
\usepackage{dcolumn}% Align table columns on decimal point
\usepackage{bm}% bold math
\begin{document}
\preprint{APS/123-QED}
\title{Non-linear corrections to the derivative of nuclear reduced cross-section at small $x$ at a future electron-ion collider}% Force line breaks with \\

\author{G.R.Boroun}%
 \email{boroun@razi.ac.ir }
 \affiliation{Department of Physics, Razi University, Kermanshah
67149, Iran}%
%\author{V. Guzey }
%\altaffiliation{vadim.a.guzey@jyu.fi}%Lines break automatically or can be forced with \\
%\affiliation{University of Jyvaskyla, Department of Physics, P.O.
%Box 35, FI-40014 University of Jyvaskyla, Finland\\
%Helsinki Institute of Physics, P.O. Box 64, FI-00014 University of Helsinki, Finland}% \textbackslash\textbackslash
\date{\today}% It is always \today, today,
\begin{abstract}
%%%%%%%%%%%%%%%%%%%%%%%%%%%%%%%%%%%%%%%%%%%%%%%%%%%%%%%
The determination of non-linear corrections to the nuclear
distribution functions due to the HIJING parametrization within
the framework of perturbative QCD, specifically the GLR-MQ
equations, is discussed. We analyze the possibility of
constraining the non-linear corrections present in distribution
functions using the inclusive observables that will be measured in
future electron-ion colliders (EIC and EICc). The results show
that  non-linear corrections play an important role in heavy
nuclear reduced cross sections at low $x$ and low $Q^2$ values. We
find that the non-linear corrections provide the correct behavior
of the extracted nuclear cross sections and that our results align
with data from the nCETQ15 parametrization group. We are currently
discussing a satisfactory description of the non-linear
corrections to the shadowing effect
at small $x$.\\

%%%%%%%%%%%%%%%%%%%%%%%%%%%%%%%%%%%%%%%%%%%%%%%%%%%%%%%
\end{abstract}
 \pacs{***}%PACS, the Physics and Astronomy
                              %Classification Scheme.
 %Use showkeys class option if keyword
\maketitle
%**********************************************************
%%%%%%%%%%%%%%%%%%%%%%%%%%%%%%%%%%%%%%%%%%%%%%%%%%%%%%%%%%%%%%%%%%%%%%%%%%%%%%%%%%%%%%%%%%%%%
\subsection{I. Introduction}

The nuclear structure can be determined from Deep Inelastic
Scattering (DIS) of leptons off nuclei across a wide range of ($x,
Q^2 $). Nuclear structure functions differ from proton structure
functions due to the shadowing effect at $x{\lesssim}0.1$,
 anti-shadowing at $0.1{\lesssim}x{\lesssim}0.3$,
the EMC effect at $0.3{\lesssim}x{\lesssim}0.7$ and Fermi motion
as $x{\rightarrow}1$. The proton structure function of the nucleus
in the leading order in the QCD-improved parton model is defined
by its parton distributions as [1]
\begin{eqnarray}
F_{2}^{p/A}(x,Q^2)=\sum_{q=u,d,s,...}e_{q}^{2}\bigg{[}xf_{q}^{p/A}(x,Q^2)+
xf_{\overline{q}}^{p/A}(x,Q^2)\bigg{]}.
\end{eqnarray}
The difference between the nuclear parton distribution functions
(nPDFs) and the parton distribution in the free proton is
determined by the ratio
\begin{eqnarray}
R_{i}^{A}(x,Q^2){\equiv}\frac{f_{i}^{p/A}(x,Q^2)}{f^{p}_{i}(x,Q^2)}.
\end{eqnarray}
The nuclear shadowing effect demonstrates that at small values of
$x$, the gluon distribution in a nucleus is less than the gluon
distribution in a nucleon. It is essential to determine the gluon
distribution of nucleons within a nucleus, especially at small $x$
values. Nuclear effects play a significant role in
$xg^{A}(x,Q^2)$, and utilizing inclusive observables can help to
constrain future electron-nucleus colliders at Brookhaven National
Laboratory (eRHIC) [2] and the Electron Ion Collider (EIC) [3].
The behavior of the nuclear gluon distribution can be determined
using the momentum sum rule. Nuclear physics with electron-nucleus
(eA) collisions can be explored at the Large Hadron electron
Collider (LHeC) [4] and the Future Circular electron-hadron
Collider (FCC-eh) [5] as proposed in Ref.[6]. The maximum energy
envisioned for electron-heavy ion runs would be achieved by
colliding $18~\mathrm{GeV}$ electrons with $110~\mathrm{GeV}$ ions
for a $\sqrt{s}=89~\mathrm{GeV}$ in the EIC Conceptual Design
Report
[7].\\
The standard evolution based on the
Dokshitzer-Gribov-Lipatov-Altarelli-Parisi (DGLAP) linear
equations provides an accurate description of QCD dynamics at
moderate to large values of the momentum fractions $x$ of the
probed parton and virtualities $Q^2{\gg}\Lambda^{2}_{QCD}$ but
needs  modification to include the effects of the resummation of
large ${\ln}(1/x)$. Gluon recombination processes, tame the growth
of parton densities towards small $x$ and lead to gluon
saturation. Non-linear evolution becomes important when the mass
number $A$ is increased or by either decreasing $x$  or some
combination of the two [6,8]. Non-linear modifications to DGLAP
evolution equations were first proposed in Refs.[9-12] where two
gluon ladders merge into a gluon or a quark-antiquark pair. The
study of non-linear corrections is indeed seful for a
comprehensive understanding of gluon recombination and saturation
[13]. The correlative interactions between gluons become important
at extremely small $x$ at fixed $Q^2$, where the probability of
recombining two gluons into one, in the leading twist
approximation, is taken to be the product of two conventional
one-gluon distributions by the following form
\begin{eqnarray}
G^{(2)}(x,Q^2)=\frac{9}{8{\pi}\mathcal{R}_{N}^{2}}[G(x,Q^2)]^2.
\end{eqnarray}
The area of a nucleon in which  gluons are populated is
characterizes by $\mathcal{R}_{N}$. This saturation tamed the
increase of gluons by relying on a Froissart-Martin bound [14].
The evolution equations of the correction terms (without the
Higher Twist (HT) terms) are given by [9-11]
\begin{eqnarray}
\frac{{\partial}}{{\partial}{\ln}Q^2}xq^{A}_{i}(x,Q^2)&=&\frac{\alpha_{s}}{2\pi}
\int_{x}^{1}\frac{dy}{y}\bigg{[}\frac{x}{y}P_{qq}\bigg{(}\frac{x}{y}
\bigg{)}yq^{A}_{i}(y,Q^2)+\frac{x}{y}P_{qg}\bigg{(}\frac{x}{y}
\bigg{)}yg^{A}(y,Q^2)\bigg{]}\nonumber\\
&&-\frac{K}{{\pi}\mathcal{R}^{2}_{A}Q^2}
\frac{2{\pi}\alpha^{2}_{s}}{N(N^2-1)}\bigg{[}\frac{4}{15}N^2-\frac{3}{5}
\bigg{]}[xg^{A}(x,Q^2)]^{2},
\end{eqnarray}
and
\begin{eqnarray}
\frac{{\partial}}{{\partial}{\ln}Q^2}xg^{A}(x,Q^2)&=&\frac{\alpha_{s}}{2\pi}
\int_{x}^{1}\frac{dy}{y}\bigg{[}\frac{x}{y}P_{gq}\bigg{(}\frac{x}{y}
\bigg{)}\sum_{i}^{2n_{f}}yq^{A}_{i}(y,Q^2)+\frac{x}{y}P_{gg}\bigg{(}\frac{x}{y}
\bigg{)}yg^{A}(y,Q^2)\bigg{]}\nonumber\\
&&-\frac{K}{{\pi}\mathcal{R}^{2}_{A}Q^2}
\frac{4{\pi^{3}}}{(N^2-1)}\bigg{(}\frac{\alpha_{s}C_{A}}{\pi}
\bigg{)}^{2}\int_{x}^{1}\frac{dy}{y}[yg^{A}(y,Q^2)]^{2},
\end{eqnarray}
where $C_{A}=N=3$, $K=\frac{9}{8}$ and
$\mathcal{R}_{A}=1.25A^{1/3}~\mathrm{fm}$ represents the nuclear
size for a nuclear target with the mass number A where represents
the gluonic hot spots inside a nucleus. The importance of the
non-linear corrections for a nuclear target\footnote{The nuclear
parton distribution functions (PDFs) scale approximately as A.}
(especially heavy nuclei) is visible, as the non-linear
terms in Eqs.(4) and (5) scale as $A^{4/3}$ [15].\\
 Adding these
contributions to the DGLAP equations yields the non-linear
Gribov-Levin-Ryskin-Mueller-Qiu (GLR-MQ) [9-10] evolution
equations for nuclei in the following forms
\begin{eqnarray}
\frac{{\partial}F^{A}_{2}(x,Q^2)}{{\partial}{\ln}Q^2}|_{\mathrm{Non-Linear}}=\frac{{\partial}F^{A}_{2}(x,Q^2)}{{\partial}{\ln}Q^2}|_{\mathrm{DGLAP}}
-2n_{f}\frac{5}{18}\frac{27\alpha^{2}_{s}}{160\mathcal{R}_{A}^2Q^2}[xg^{A}(x,Q^2)]^2
\end{eqnarray}
and
\begin{eqnarray}
\frac{{\partial}xg^{A}(x,Q^2)}{{\partial}{\ln}Q^2}|_{\mathrm{Non-Linear}}=\frac{{\partial}xg^{A}(x,Q^2)}{{\partial}{\ln}Q^2}|_{\mathrm{DGLAP}}
-\frac{81\alpha^{2}_{s}}{16\mathcal{R}_{A}^2Q^2}\int_{x}^{1}\frac{dy}{y}[yg^A(y,Q^2)]^2,
\end{eqnarray}
where the non-linear term tames the growth of the distribution
functions at small $x$ and leads to their suppression [15-17].
Here $xg^{A}(x,Q^2)$ is the gluon distribution function of nuclei
and
$F_{2}^{A}(x,Q^2)=\sum{e_{i}^{2}}[xq_{i}^{A}(x,Q^2)+x\overline{q}_{i}^{A}(x,Q^2)]$
where $e_{i}$ is the electric charge of the $i$-quark or antiquark
and $q_{i}^{A}(x,Q^2)$ is the number density of the $i$-quarks in
the nucleus.\\
In this paper, we examine the reduced cross sections for light and
heavy nuclei at the EIC center-of -mass (COM) energy. We then
explore the recombination of the derivative of the reduced cross
section into $\ln{Q^2}$ across a wide range of light and
heavy nuclei.\\

%%%%%%%%%%%%%%%%%%%%%%%%%%%%%%%%%%%%%%%%%%%%%%%%%%%%%%%%%%%%%%%%%%%%%%%%%%%%%%%%%%%%%%
\subsection{II. Deep inelastic lepton-nucleus scattering}

The double differential cross section for deep inelastic
scattering (DIS) of an electron-nucleus from an unpolarized
nucleus in the one photon exchange approximation has the following
form
\begin{eqnarray}
\frac{d^{2}\sigma^{A}}{dxdQ^2}=\frac{2{\pi}\alpha^2}{xQ^2}Y_{+}\sigma_{r}^{A}(x,Q^2).
\end{eqnarray}
Here, $Y_{+}=1+(1-y)^2$ and $y$ represents the inelasticity. The
nuclear reduced cross section $\sigma_{r}^{A}$ can be standardly
defined using the structure functions $F_{2}^{A}$ and $F_{L}^{A}$
as follows [6]
\begin{eqnarray}
\sigma_{r}^{A}(x,Q^2)=F_{2}^{A}(x,Q^2)-\frac{y^2}{Y_{+}}F_{L}^{A}(x,Q^2).
\end{eqnarray}
The longitudinal structure function in nuclear deep inelastic
scattering (nDIS) is an observable that can be used to unfold the
gluon distribution [18]. nQCD provides the Altarelli-Martinelli
equation [19] in the following form
\begin{eqnarray}
F_{L}^{A}(x,Q^2)&=&\frac{\alpha_{s}(Q^2)}{2\pi}x^2\int_{x}^{1}\frac{dz}{z^3}\bigg{[}
\frac{8}{3}F_{2}^{A}(z,Q^2)+4\sum{e_{q}^{2}}(1-\frac{x}{z})zg^{A}(z,Q^2)\bigg{]}.
\end{eqnarray}
The scheme-independent coefficient functions for the longitudinal
structure function can be found in Ref.[20]. The nuclear effects
for the eA scattering can be defined by the ratio of distribution
functions as
\begin{eqnarray}
R_{F_{2}}^{A}(x)=\frac{F_{2}^{A}(x,Q^2)}{AF_{2}(x,Q^2)},
\end{eqnarray}
and
\begin{eqnarray}
R_{g}^{A}(x)=\frac{xg^{A}(x,Q^2)}{Axg(x,Q^2)},
\end{eqnarray}
where $xg(x,Q^2)$ and $F_{2}(x,Q^2)$ are respectively the gluon
distribution and the structure function of a free nucleon.\\
 The expression for $\sigma_{r}^{A}$ can be rewritten as a
function of the structure function $F^{A}_{2}(x,Q^2)$ and the
gluon distribution $xg^{A}(x,Q^2)$ of nuclei in the following form
\begin{eqnarray}
\sigma_{r}^{A}(x,Q^2)&=&F^{A}_{2}(x,Q^2)-\frac{y^2\alpha_{s}(Q^2)}{2{\pi}Y_{+}}x^2\int_{x}^{1}\frac{dz}{z^3}
\bigg{[}\frac{8}{3}F^{A}_{2}(z,Q^2)
+4\sum{e_{q}^{2}}(1-\frac{x}{z})zg^{A}(z,Q^2)\bigg{]}.
\end{eqnarray}
Nuclear effects are shown in the ratio of distribution functions.
Parameterizations of the nuclear parton distribution functions
have been proposed by some groups in Refs.[21-25] and extended in
recent years in Refs.[26-28, 8]. The HIJING2.0 [26]
parametrization which is in good agreement with the ALICE
experiment at LHC energies, provides a more stringent constraint
on gluon shadowing due to the impact parameter dependence of the
shadowing as reported in Refs.[27-28] for light and heavy nuclei
\begin{eqnarray}
R^{A}_{F_{2}}(x)&=& 1+1.19(\ln A)^{1/6}(x^3-1.2x^2+0.21x)
-s_{q}\frac{5}{3}(1-b^2/\mathcal{R}_{A}^{2})(A^{1/3}-1)^{0.6}(1-3.5\sqrt{x})\exp(-x^2/0.01),
\end{eqnarray}
and
\begin{eqnarray}
R^{A}_{g}(x)&=& 1+1.19(\ln A)^{1/6}(x^3-1.2x^2+0.21x)
-s_{g}\frac{5}{3}(1-b^2/\mathcal{R}_{A}^{2})(A^{1/3}-1)^{0.6}(1-1.5{x}^{0.35})\exp(-x^2/0.004),
\end{eqnarray}
where  $s_{q} = 0.1$, $s_{g}= 0.22-0.23$. In this case, the impact
parameter $b$ is chosen as central with $b=0$ for light nuclei
and peripheral with $b=5~\mathrm{fm}$ for heavy nuclei [27-28].\\
The non-linear correction to the derivative of the nuclear
structure function divided  by A (according to Eq.(6)) is defined
as follows
\begin{eqnarray}
\frac{1}{A}\frac{\partial{\Delta}F_{2}^{A}(x,Q^2)}{\partial{\ln}Q^2}=2n_{f}\frac{5}{18}\frac{27\alpha_{s}^{2}(Q^2)}{160\mathcal{R}_{A}^{2}Q^2}A[R^{A}_{g}(x)xg(x,Q^2)]^2.
\end{eqnarray}
This equation defines the magnitude  of the non-linear corrections
as
$${\Delta}F_{2}^{A}(x,Q^2)=F_{2}^{A}(x,Q^2)|_{\mathrm{DGLAP}}-F_{2}^{A}(x,Q^2)|_{\mathrm{Non-Linear}}.$$
Non-linear corrections can be determined from the inclusive
nuclear cross section in the low $x$ and $Q^2$ region. This
behavior can be utilized in a derivative method in an EIC based on
the cross section derivative. The derivative of the reduced cross
section for nuclei is expressed as
\begin{eqnarray}
\frac{{\partial}\sigma_{r}^{A}(x,Q^2)}{{\partial}\ln{Q^2}}|_{y=cte}&=&\frac{{\partial}F^{A}_{2}(x,Q^2)}{{\partial}\ln{Q^2}}
-\frac{y^2\alpha_{s}(Q^2)}{2{\pi}Y_{+}}x^2
\int_{x}^{1}\frac{dz}{z^3}\bigg{[}\frac{8}{3}\bigg{\{}\frac{{\partial}F^{A}_{2}(z,Q^2)}{{\partial}\ln{Q^2}}
+\frac{{\partial}{\ln}\alpha_{s}(Q^2)}{{\partial}\ln{Q^2}}F_{2}^{A}(z,Q^2)\bigg{\}}\nonumber\\
&&+4\sum{e_{q}^{2}}(1-\frac{x}{z})\bigg{\{}\frac{{\partial}zg^{A}(z,Q^2)}{{\partial}\ln{Q^2}}
+\frac{{\partial}{\ln}\alpha_{s}(Q^2)}{{\partial}\ln{Q^2}}zg^{A}(z,Q^2)\bigg{\}}\bigg{]}.
\end{eqnarray}
Gluon recombination alters the behavior of the parton densities
and introduces nonlinear effects. Consequently, the derivative of
the nuclear reduced cross section is adjusted due to these
nonlinear effects:
\begin{eqnarray}
\frac{1}{A}\frac{{\partial}\sigma_{r}^{A}(x,Q^2)}{{\partial}\ln{Q^2}}|_{\mathrm{Non-Linear}}=\frac{1}{A}\frac{{\partial}\sigma_{r}^{A}(x,Q^2)}{{\partial}\ln{Q^2}}|_{\mathrm{Eq.17}}
-2n_{f}\frac{5}{18}\frac{27\alpha_{s}^{2}(Q^2)}{160\mathcal{R}_{A}^2Q^2}
A[R_{g}^{A}(x)xg(x,Q^2)]^2-\mathcal{O}(\alpha_{s}^{3}),~~~
\end{eqnarray}
where $\mathcal{O}(\alpha_{s}^{3})$ represents the non-linear
effects to the derivative of the longitudinal structure function
of nuclei as
\begin{eqnarray}
\mathcal{O}(\alpha_{s}^{3})&=&A\frac{y^2\alpha^{3}_{s}(Q^2)}{2{\pi}Y_{+}}\frac{x^2}{\mathcal{R}_{A}^2Q^2}\int_{x}^{1}\frac{dz}{z^3}
\bigg{[}\frac{2n_{f}}{8} [R_{g}^{A}(z)zg(z,Q^2)]^2
+\frac{81}{4}\sum{e_{q}^{2}}(1-\frac{x}{z})\int_{z}^{1}\frac{d\xi}{\xi}[R_{g}^{A}(\xi){\xi}g(\xi,Q^2)]^2\bigg{]},~~~~
\end{eqnarray}
where at moderate inelasticity we observe that the term
$\mathcal{O}(\alpha_{s}^{3})$ is very small across a wide range of
$x$, therefore
\begin{eqnarray}
\mathcal{O}(\alpha_{s}^{3}){\sim}0.
\end{eqnarray}
%\begin{figure}
%\includegraphics[width=0.6\textwidth]{Fig1}
%\caption{Results of order $\alpha_{s}^{3}$ in Eq.(16) are shown as
%a function of $x$ in a wide range of $y$ at $Q^2=5$  and
%$10~\mathrm{GeV}^2$ for the light nucleus of C-12
% and the heavy nucleus
%  of Pb-208 due to the DL [25] and Block et.al., parametrization [26]. }\label{Fig1}
%\end{figure}
In conclusion, we can safely ignore this term and simplify Eq.(18)
for the derivative of the reduced cross section of nuclei to the
following form
\begin{eqnarray}
\frac{1}{A}\frac{{\partial}\sigma_{r}^{A}(x,Q^2)}{{\partial}\ln{Q^2}}|_{\mathrm{Non-Linear}}{\simeq}\frac{1}{A}\frac{{\partial}\sigma_{r}^{A}(x,Q^2)}{{\partial}\ln{Q^2}}|_{\mathrm{Eq.17}}
-2n_{f}\frac{5}{18}\frac{27\alpha_{s}^{2}(Q^2)}{160\mathcal{R}^2Q^2}
A[R_{g}^{A}(x)xg(x,Q^2)]^2,~~~
\end{eqnarray}
which is similar to the GLR-MQ evolution equations. Indeed, the
effect of the non-linear corrections to the derivative of the
reduced cross section of nuclei divided by A is defined as
\begin{eqnarray}
\frac{1}{A}\frac{{\partial}{\Delta}\sigma_{r}^{A}(x,Q^2)}{{\partial}\ln{Q^2}}=2n_{f}\frac{5}{18}\frac{27\alpha_{s}^{2}(Q^2)}{160\mathcal{R}^2Q^2}A[R_{g}^{A}(x)xg(x,Q^2)]^2.~~~~
\end{eqnarray}
This is similar to the derivative of the structure functions of
nuclei divided by A as
\begin{eqnarray}
\frac{1}{A}\frac{{\partial}{\Delta}\sigma_{r}^{A}(x,Q^2)}{{\partial}\ln{Q^2}}|_{y=cte}{\simeq}\frac{1}{A}\frac{\partial{\Delta}F_{2}^{A}(x,Q^2)}{\partial{\ln}Q^2}.
\end{eqnarray}
In the following, we consider the non-linear effects on the
reduced cross section of nuclei divided by A, based on shadowing
effects. The non-linear corrections at the initial scale $Q_{0}^2$
are adjusted by applying shadowing corrections [29] for
$x<x_{0}{\equiv}10^{-2}$ through the nuclear parton distribution
functions as
\begin{eqnarray}
xg^{A}(x,Q_{0}^2){\rightarrow}xg^{A}(x,Q_{0}^2)\zeta^{A}(x,x_{0},Q_{0}^2)\nonumber\\
\mathrm{and}\hspace{3.5cm}\nonumber\\
xq^{A}_{s}(x,Q_{0}^2){\rightarrow}xq^{A}_{s}(x,Q_{0}^2)\zeta^{A}(x,x_{0},Q_{0}^2),
\end{eqnarray}
where
\begin{eqnarray}
\zeta^{A}(x,x_{0},Q_{0}^2)&=&\Big{\{}1+\theta(x_{0}-x) \Big{[}
xg^{A}(x,Q_{0}^2)-xg^{A}(x_{0},Q_{0}^2)
\Big{]}/xg^{A}_{\mathrm{sat}}(x,Q_{0}^{2})\Big{\}}^{-1},
\end{eqnarray}
with
\begin{eqnarray}
xg^{A}_{\mathrm{sat}}(x,Q^{2})=\frac{16{\mathcal{R}_{A}}^{2}Q^2}{27{\pi}\alpha_{s}(Q^2)},
\end{eqnarray}
where $g^{A}_{\mathrm{sat}}$ is the value of the gluon that would
saturate the unitarity limit in the leading shadowing
approximation in nuclei. The non-linear corrections to the reduced
cross sections of nuclei are defined by the following form
\begin{eqnarray}
\sigma_{r}^{A}(x,Q^2)|_{\mathrm{Non-Linear}}&=&F^{A}_{2}(x,Q^2)|_{\mathrm{Non-Linear}}-\frac{y^2\alpha_{s}(Q^2)}{2{\pi}Y_{+}}x^2\int_{x}^{1}\frac{dz}{z^3}
\bigg{[}\frac{8}{3}F^{A}_{2}(z,Q^2)|_{\mathrm{Non-Linear}}\nonumber\\
&&+4\sum{e_{q}^{2}}(1-\frac{x}{z})zg^{A}(z,Q^2)|_{\mathrm{Non-Linear}}\bigg{]},~~~~
\end{eqnarray}
where
\begin{eqnarray}
F_{2}^{A}(x,Q^2)|_{\mathrm{Non-Linear}}=F^{A}_{2}(x,Q^2)+F^{A}_{2}(x,Q_{0}^2)(\zeta^{A}(x,x_{0},Q_{0}^2)-1)
-2n_{f}\frac{5}{18}\frac{27}{160\mathcal{R}_{A}^2}\int_{Q_{0}^{2}}^{Q^{2}}\frac{\alpha_{s}^{2}(q^2)}{q^2}[xg^{A}(x,q^2)]^2d{\ln}q^2,~~~
\end{eqnarray}
and
\begin{eqnarray}
xg^{A}(x,Q^2)|_{\mathrm{Non-Linear}}=xg^{A}(x,Q^2)+xg^{A}(x,Q_{0}^2)(\zeta^{A}(x,x_{0},Q_{0}^2)-1)
-\frac{81}{16\mathcal{R}_{A}^2}\int_{Q_{0}^{2}}^{Q^{2}}\frac{\alpha_{s}^{2}(q^2)}{q^2}\int_{x}^{1}\frac{dy}{y}[yg^A(y,Q^2)]^2d{\ln}q^2.~~~
\end{eqnarray}
Therefore, we find that the derivative of the reduced cross
section divided by A, due to the non-linear corrections, is
defined by the following form
\begin{eqnarray}
\frac{1}{A}{\Delta}\sigma^{A}_{r}&=& 2n_{f}\frac{5}{18}
\frac{27}{160\mathcal{R}_{A}^2}A\int_{Q_{0}^{2}}^{Q^2}\frac{\alpha_{s}^{2}(q^2)}{q^2}[R_{g}^{A}(x)xg(x,q^2)]^2d{\ln}q^2-R_{F_{2}}^{A}(x)F_{2}(x,Q_{0}^{2})(\zeta^{A}(x,x_{0},Q_{0}^{2})-1)\nonumber\\
&&+\frac{y^2\alpha_{s}(Q^2)}{2{\pi}Y_{+}}x^2\int_{x}^{1}\frac{dz}{z^3}(\zeta^{A}(z,x_{0},Q_{0}^{2})-1)
\bigg{[}\frac{8}{3}R_{F_{2}}^{A}(z)F_{2}(z,Q_{0}^2)+4\sum{e_{q}^{2}}(1-\frac{x}{z})R_{g}^{A}(z)
zg(z,Q_{0}^2)\bigg{]}\nonumber\\
&&-\frac{A}{8\mathcal{R}_{A}^2}\frac{y^2\alpha_{s}(Q^2)}{{\pi}Y_{+}}x^2\int_{x}^{1}\frac{dz}{z^3}\bigg{\{}
n_{f}\int_{Q_{0}^{2}}^{Q^2}\frac{\alpha_{s}^{2}(q^2)}{q^2}[R_{g}^{A}(z)zg(z,q^2)]^2d{\ln}q^2\nonumber\\
&&+81\sum{e_{q}^{2}}(1-\frac{x}{z})\int_{Q_{0}^{2}}^{Q^2}\frac{\alpha_{s}^{2}(q^2)}{q^2}\int_{z}^{1}\frac{d\xi}{\xi}[R_{g}^{A}(\xi){\xi}g(\xi,q^2)]^2d{\ln}q^2\bigg{\}}.
\end{eqnarray}
In the following analysis, the gluon distribution and the proton
structure functions are defined using the Donnachie-Landshoff (DL)
[30] and Block et al. [31-32]
methods (please see the Appendix).\\
Therefore, the non-linear correction to the nuclear shadowing
effect, which is associated with the modification of the target
gluon recombination, is defined as
\begin{eqnarray}
\frac{{\partial}{\Delta}\sigma_{r}^{A}(x,Q^2)}{A{\partial}{\Delta}\sigma_{r}(x,Q^2)}|_{y=cte}{\simeq}
\frac{\partial{\Delta}F_{2}^{A}(x,Q^2)}{A\partial{\Delta}F_{2}(x,Q^2)}=A(R_{g}^{A})^2\frac{\mathcal{R}^{2}}{\mathcal{R}_{A}^2}.
\end{eqnarray}
This equation (i.e., Eq.(31)) predicts the modification of the
magnetite due to non-linear corrections from gluon recombination
in  nucleons and nuclei. This new phenomenon could be a key factor
in the Color Glass Condensate (CGC) [34] theory. The non-linear
correction to shadowing in nuclei (i.e., Eq.(31)) is examined by
comparing the non-linear corrections to the structure functions
per nucleon for various nuclei\footnote{The nuclear ratio in the
presence of saturation , considering geometric scaling, is
discussed in Ref.[35] with a simple parameterization for the
unintegrated gluon distribution based on the asymptotic solutions
of the Balitsky-Kovchegov (BK) equation [36]. }.

\subsection{III. Results and Conclusions}

The QCD parameter $\Lambda$ is extracted from the running coupling
$\alpha_{s}(Q^2)$, where
$\Lambda_{\mathrm{QCD}}=0.12~\mathrm{GeV}$ yields
$\alpha_{s}(M_{z}^2)=0.118$ for the one-loop coupling, with the
number of active flavors being $n_{f}=4$. The behavior of the
reduced cross section and the non-linear corrections to the
derivative of the nuclear reduced cross section are determined for
the light nucleus of C-12
 and the heavy nucleus of Pb-208 at the hot-spot point
 $\mathcal{R}_{A}=1.25A^{1/3}~\mathrm{fm}$ in Figs.1-6. The results are
 reported for the kinematic range relevant for the EIC
 ($\sqrt{s}=89~\mathrm{GeV}$ and $y$ less than roughly 1), which is evident in the fact that $x$ values are
  plotted down to $~0.65{\times}10^{-3}$ (or $y{\simeq}0.97$) for $Q^2=5~\mathrm{GeV}^2$ ($x=Q^2/ys$)
  and $~0.13{\times}10^{-2}$ for $Q^2=10~\mathrm{GeV}^2$.\\
For comparison with the nCTEQ15 nPDFs [33] results, we calculate
the expression
$\frac{1}{A}\frac{\partial}{\partial{\ln}Q^2}{\Delta}F_{2}^{A}(x,Q^2)$,
which quantifies the effect of non-linear corrections. In Fig.1,
the non-linear corrections to
$\frac{1}{A}\frac{\partial}{\partial{\ln}Q^2}{\Delta}F_{2}^{A}(x,Q^2)$
for the heavy nucleus of Pb-208  are plotted as a function of the
momentum fraction $x$ at $Q^2=5~\mathrm{GeV}^2$. The results are
determined with respect to the DL (square-purple) [30] and Block
et al (circle-brown) [31-32] methods. They are also compared to
the nCTEQ15 parametrization with uncertainties at corresponding
values of $Q^2$ represented by the solid curve (red,
$\mathcal{R}=2~\mathrm{GeV}^{-1}$), the dashed curve (blue,
$\mathcal{R}=5~\mathrm{GeV}^{-1}$) and the dashed-dot curve
(black, $\mathcal{R}=1.25A^{1/3}~\mathrm{fm}$). These results  are
comparable to the nCTEQ15 parametrization at
$\mathcal{R}=1.25A^{1/3}~\mathrm{fm}$. The nCTEQ15 parametrization
results in a wide range of $x$ are flat, while our results
increase as $x$ decreases. This difference is due to the behavior
of  the DL and Block et al gluon distribution functions. With an
increase in $Q^2$ values, the nCTEQ15 parametrization results
increase as $x$ decreases. In Fig.2, we show this behavior for the
heavy nucleus of Pb-208 at $Q^2=10~\mathrm{GeV}^2$. Our results
are comparable to the nCTEQ15 parametrization results accompanied
by uncertainties at small $x$ values. The difference between the
results with the nCTEQ15 parametrization results at
$\mathcal{R}=1.25A^{1/3}~\mathrm{fm}$
at moderate $x$ values is due to the gluon dominance solely in our results.\\
\begin{figure}
\includegraphics[width=0.6\textwidth]{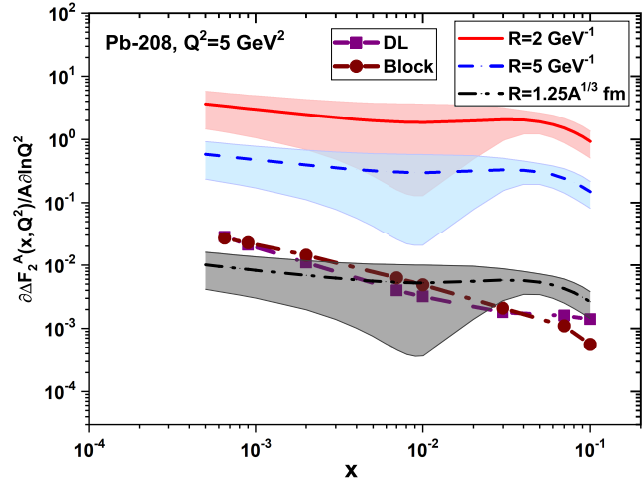}
\caption{The non-linear corrections to
$\frac{1}{A}\frac{\partial}{\partial{\ln}Q^2}{\Delta}F_{2}^{A}(x,Q^2)$
for the heavy nucleus of Pb-208 are shown as a function of the
momentum fraction $x$ at $Q^2=5~\mathrm{GeV}^2$ at
$\mathcal{R}_{A}=1.25A^{1/3}~\mathrm{fm}$. These results are
determined by the DL (square-purple) [30] and Block et al
(circle-brown) [31-32] gluon distributions and compared with the
nCETQ15 parametrization [33] results at
$\mathcal{R}=2~\mathrm{GeV}^{-1}$ (solid curve-red),
$\mathcal{R}=5~\mathrm{GeV}^{-1}$ (dashed curve- blue ) and
$\mathcal{R}=1.25A^{1/3}~\mathrm{fm}$ (dashed- dot curve- black)
with uncertainties. }\label{Fig1}
\end{figure}
\begin{figure}
\includegraphics[width=0.55\textwidth]{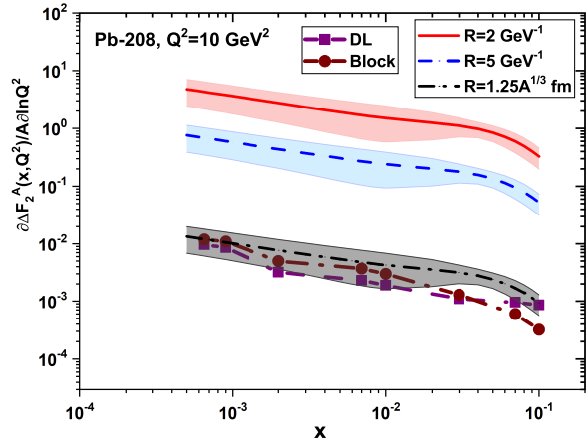}
\caption{The same as Fig.1 for Pb-208 at
$Q^2=10~\mathrm{GeV}^2$.}\label{Fig2}
\end{figure}
In the following, we present the non-linear corrections to the
derivative of the
  nuclear reduced cross section into ${\ln}Q^2$ divided by A, $\frac{1}{A}\frac{{\partial}}{{\partial}{\ln}Q^2}\Delta\sigma_{r}^A(x,Q^2)$, for the light nucleus of C-12
   and the heavy nucleus
  of Pb-208 according
  to the EIC COM energy  at the fixed value of the
  inelasticity $y$ (for $y=0.2$ and $y=0.6$). In Figs.3 and 4, these results are obtained with respect to the DL [30] and  the Block et
al., methods [31-32]  are presented respectively.  We observe that
these
  non-linear corrections are visible at high inelasticity and
  small $Q^2$ values
  for the light nucleus of C-12
   and the heavy nucleus
  of Pb-208.\\
\begin{figure}
\includegraphics[width=0.55\textwidth]{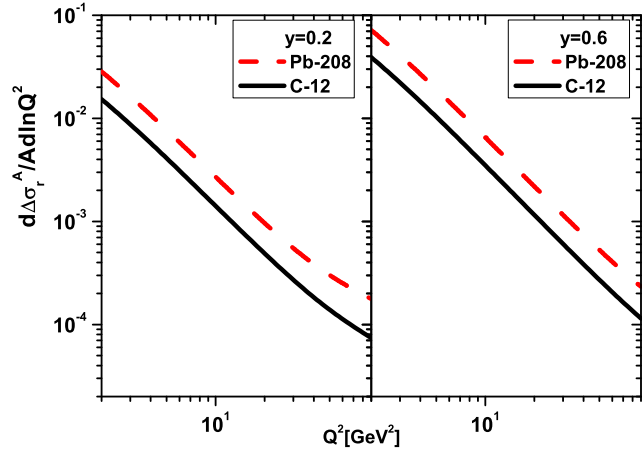}
\caption{Results of
$\frac{1}{A}\frac{{\partial}}{{\partial}{\ln}Q^2}\Delta\sigma_{r}^A(x,Q^2)$
are shown as a function of $Q^2$ at $y=0.2$ (left) and $y=0.6$
(right) for the light nucleus of C-12 (black-solid curve)
   and the heavy nucleus
  of Pb-208 (red-dashed curve) due to the DL method [30]. }\label{Fig3}
\end{figure}
\begin{figure}
\includegraphics[width=0.55\textwidth]{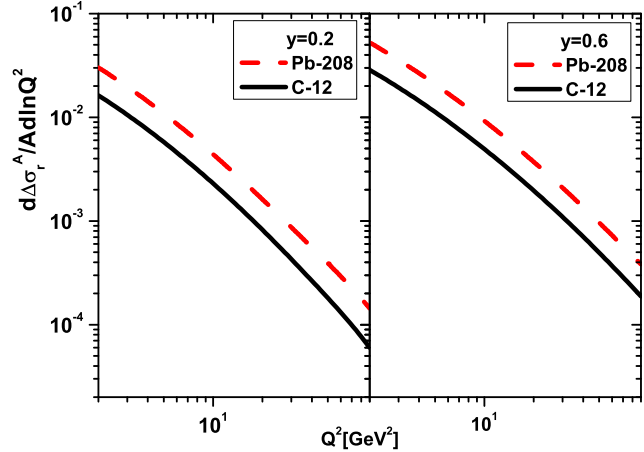}
\caption{The same as Fig.3 due to the Block et al., method
[31-32].}\label{Fig4}
\end{figure}
In Fig.5, we present results of our numerical studies to the
non-linear corrections of the nuclear reduced cross section
divided by A, $\frac{1}{A}\Delta\sigma_{r}^A(x,Q^2)$, for the
light nucleus of C-12
   and the heavy nucleus
  of Pb-208 according
  to the EIC COM energy  at the fixed value of   $Q^{2}$ (for $Q^{2}=5$ and $10~\mathrm{GeV}^2$). In Fig.5, these results are presented with respect to
  the gluon distribution of the DL [30] method. We observe that these
  non-linear corrections are visible at low $x$ values
  for the light nucleus of C-12
   and the heavy nucleus
  of Pb-208.\\
\begin{figure}
\includegraphics[width=0.55\textwidth]{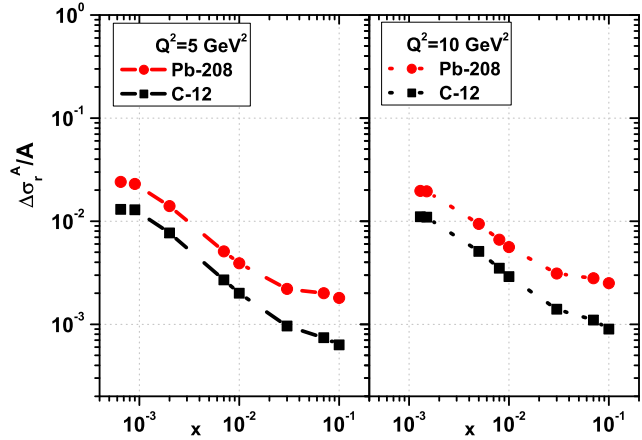}
\caption{Results of $\frac{1}{A}\Delta\sigma_{r}^A(x,Q^2)$ are
shown as a function of $Q^2$ at $Q^2=5~\mathrm{GeV}^2$ (left) and
$Q^2=5~\mathrm{GeV}^2$ (right) for the light nucleus of C-12
(black-square points)
   and the heavy nucleus
  of Pb-208 (red-circle points) due to the DL method [30].}\label{Fig5}
\end{figure}
\begin{figure}
\includegraphics[width=0.55\textwidth]{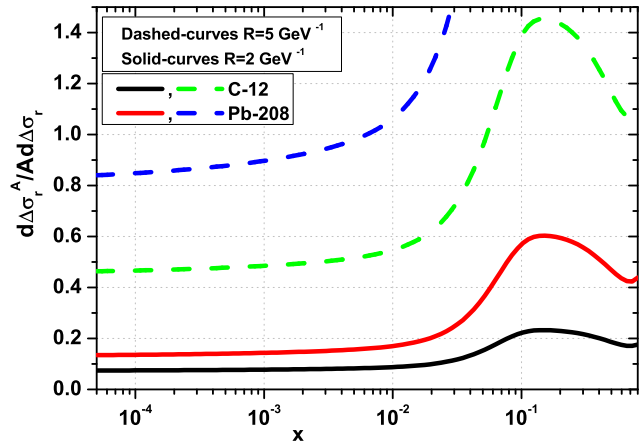}
\caption{Ratio
$\frac{{\partial}{\Delta}\sigma_{r}^{A}(x,Q^2)}{A{\partial}{\Delta}\sigma_{r}(x,Q^2)}|_{y=cte}{\simeq}
\frac{\partial{\Delta}F_{2}^{A}(x,Q^2)}{A\partial{\Delta}F_{2}(x,Q^2)}=A(R_{g}^{A})^2\frac{\mathcal{R}^{2}}{\mathcal{R}_{A}^2}$
as a function of Bjorken-$x$ for the light nucleus of C-12 and the
heavy nucleus Pb-208 with
$\mathcal{R}_{A}=1.25A^{1/3}~\mathrm{fm}$. $R_{g}^{A}$ is
predicted by the HIJING parametrization. Solid curves represent
the ratio for the light nucleus of C-12 (Black) and heavy nucleus
Pb-208 (Red) with $\mathcal{R}=2~\mathrm{GeV}^{-1}$ and dashed
curves represent the ratio for the light nucleus of C-12 (Green)
and heavy nucleus Pb-208 (Blue) with
$\mathcal{R}=5~\mathrm{GeV}^{-1}$. }\label{Fig6}
\end{figure}
In Fig.6, the ratio
$\frac{{\partial}{\Delta}\sigma_{r}^{A}(x,Q^2)}{A{\partial}{\Delta}
\sigma_{r}(x,Q^2)}|_{y=cte}$ is shown, which is approximately
equal to
$\frac{\partial{\Delta}F_{2}^{A}(x,Q^2)}{A\partial{\Delta}F_{2}(x,Q^2)}=A(R_{g}^{A})^2\frac{\mathcal{R}^{2}}{\mathcal{R}_{A}^2}$
. This comparison is made between for the light nucleus of C-12
and the heavy nucleus Pb-208 as a function of Bjorken-$x$ to
determine the non-linear correction to the saturation effect in
nuclei. It is evident that the magnitude of shadowing due to the
non-linear corrections is well-defined. This indicates that
shadowing effects resulting from the non-linear corrections can be
readily constrained at the EIC for $x<10^{-2}$, which strongly
depends on the proton hot-spot point and the mass number A.
Therefore, by measuring
$A(R_{g}^{A})^2\frac{\mathcal{R}^{2}}{\mathcal{R}_{A}^2}$, it is
 possible to determine the existence and magnitude of the
non-linear correction to the shadowing effect in the GLR-MQ
evolution equation, which is a crucial quantity for probing
nuclear effects and QCD dynamics at small-$x$.

In conclusion, we have examined the non-linear corrections for
various values of $x$ and found that the shadowing effect in the
GLR-MQ equations increases for heavy nuclei. We have analyzed the
behavior of the logarithmic slopes of the nuclear structure
function and the nuclear reduced cross section in the kinematic
region of future electron-ion colliders (LHeC, FCC-eh and EIC).
These results, using the HIJING parametrization, suggest a
decrease in the nuclear cross section in future electron-ion
colliders. The growth of the reduced cross section divided by A
for the heavy nucleus Pb-208 and the light nucleus C-12 at small
$x$ is controlled at low values of $Q^2$ at the hot spot point
$\mathcal{R}_{A}=1.25A^{1/3}~\mathrm{fm}$. This gluonic hot spot
structure in the nucleus is significant for EIC collisions. The
magnitude of
$\frac{1}{A}\frac{{\partial}}{{\partial}{\ln}Q^2}\Delta\sigma_{r}^A(x,Q^2)$
and $\frac{1}{A}\Delta\sigma_{r}^A(x,Q^2)$ increases as $x$
decreases  and the atomic number A increases. The behavior of
$\frac{1}{A}\frac{\partial}{\partial{\ln}Q^2}{\Delta}F_{2}^{A}(x,Q^2)$
for the heavy nucleus Pb-208 is compared to the results of the
nCTEQ15 parametrization at $Q^{2}=5$ and $10~\mathrm{GeV}^2$. Our
analysis indicates that the non-linear corrections are quite
significant at $x{\sim}10^{-3}$ and high inelasticity according to
the EIC COM. These results demonstrate that the inclusive
observables are sensitive to the non-linear corrections. Drawing
firm conclusions about the QCD dynamics from the nuclear reduced
cross sections in the kinematic range of future electron-ion
experiments is possible.
%%%%%%%%%%%%%%%%%%%%%%%%%%%%%%%%%%%%%%%%%%%%%%%%%%%%%%%%%%%
\subsection{ACKNOWLEDGMENTS}

The author is grateful to Razi University for the financial
support provided for this project. Additionally, the author would
like to express thanks to Professor
Vadim Guzey for his helpful comments and invaluable support.\\

%%%%%%%%%%%%%%%%%%%%%%
%%%%%%%%%%%%%%%%%%%%%%%%%%%%%%%%%%%%%%%%%%%%%%%%%%%%%%%%%%%%%%%%%%%%%%%%
%\begin{table}[h]
%\caption{ The coefficients [5] at low $x$ for
%$0.15~\mathrm{GeV}^{2}<Q^{2}<3000~\mathrm{GeV}^{2}$.}
%\begin{tabular} {cccc}
%\toprule \\  \multicolumn{2}{c}{parameters \quad \quad \quad ~~~~~~~~~~~~~~~~value}    \\ &&&\\ \hline \\ &&&\\

%$a_{00}$ & \quad $ 0.255 \pm 0.016$ & &\\

%$a_{01}$& \quad  $1.475\times 10^{-1}~\pm 3.025\times10^{-2}$ & &\\&&&\\

 % $a_{10} $  &   \quad  $8.205\times 10^{-4}~~  \pm  4.62\times10^{-4} $  \\

%  $a_{11} $  &   \quad   $-5.148\times 10^{-2}\pm 8.19\times10^{-3}$  \\

%  $a_{12}$   &    \quad  $-4.725\times 10^{-3}\pm 1.01\times10^{-3}$   \\  &&&\\

% $a_{20}$   &   \quad   $2.217\times 10^{-3}\pm 1.42\times10^{-4} $ \\

% $a_{21}$   &   \quad   $1.244\times 10^{-2}\pm 8.56\times10^{-4}$  \\

% $a_{22}$    &    \quad  $5.958\times 10^{-4}\pm 2.32\times10^{-4} $ \\ &&& \\

%$n$& \quad  $11.49\pm 0.99$ & &\\

%$\lambda$& \quad  $2.430~\pm 0.153$ & &\\

%$M^{2}$ & \quad $0.753 \pm 0.068~ \mathrm{GeV}^{2}$ & &\\

%$\mu^2$ & \quad $ 2.82 \pm 0.290~ \mathrm{GeV}^{2}$ & &\\

%$\chi^{2}(\mathrm{goodness~ of~ fit})$ &  \quad  $0.95$ & &\\
%\hline

%\end{tabular}
%\end{table}

%%%%%%%%%%%%%%%%%%%%%%%%%%%%%%%%%%%%%%%%%%%%%%
%%%%%%%%%%%%%%%%%%%%%%%%%%%%%%%%%%%%%%%%%%%%%%
\section{Appendix}

For the reduced cross-section of nuclei, as previously mentioned,
we require the nucleon distribution functions to be in terms of
the variables $x$ and $Q^2$. The gluon distribution function and
the proton structure function are initially parametrized by
Donnachie-Landshoff [30] for the deep inelastic structure function
in electromagnetic scattering with protons. The structure function
$F_{2}(x,Q^{2})$ parametrized by Donnachie-Landshoff, at large
$W=\sqrt{s}$, is expressed as
\begin{eqnarray}
F_{2}(x,Q^{2}){\sim}f_{0}(Q^2)x^{-\epsilon_{0}},
\end{eqnarray}
where
\begin{eqnarray}
f_{0}(Q^2)=X_{0}(Q^2)^{1+\epsilon_{0}}(1+Q^2/Q_{0}^2)^{-1-\frac{1}{2}\epsilon_{0}}.
\end{eqnarray}
The proton structure function data indicates the presence of a
hard pomeron, with an intercept of $1+\epsilon_{0}$ at small $x$.
The fitted results to the ZEUS and H1 data in the range $x<0.001$
and $0.045{\leq}Q^2{\leq}35~\mathrm{GeV}^2$ are provided as
follows[30]:
$$
X_{0}=0.00146,~~Q_{0}^{2}=9.11~\mathrm{GeV}^2,~~\epsilon_{0}=0.437
$$
The charmed quark component $F_{2}^{c}$ of $F_{2}$ is
predominantly influenced by hard pomeron exchange at small $x$.
Therefore, a numerical fit to the solution of the DGLAP evolution
for the gluon distribution at small $x$ is defined as [30]:
\begin{eqnarray}
xg(x,Q^2){\sim}0.95(Q^2)^{1+\epsilon_{0}}(1+Q^2/0.5)^{-1-\frac{1}{2}\epsilon_{0}}x^{-\epsilon_{0}}.
\end{eqnarray}
The Donnachie-Landshoff parametrization of the distribution
functions has a limited range of applicability. In Ref.[31], the
authors have presented a parametrization of $F_{2}$ that applies
to large and small $Q^2$ using the proposed Froissart-bound. This
parameterization provides an excellent fit to all available ZEUS
and H1 data across a wide range of $x$ and $Q^2$. The explicit
expression for the BBT parametrization [32] is as follows:
\begin{eqnarray}
F_{2}(x,Q^2)&=&(1-x)\bigg{[}\frac{F_{P}}{1-x_{p}}+(a_{0}+\sum_{m=1}^{2}a_{m}\nonumber\\
&&{\times}\ln^{m}(Q^2))\ln\bigg{[} \frac{x_{P}(1-x)}{x(1-x_{P})}\\
&&+(b_{0}+\sum_{m=1}^{2}b_{m}\ln^{m}(Q^2))\ln^{2}\bigg{[}
\frac{x_{P}(1-x)}{x(1-x_{P})}\bigg{]},\nonumber
\end{eqnarray}
with $F_{P}=0.41$ and $x_{P}=0.09$ (The other coefficients are
shown in Table I). In Ref.[31], the authors have derived a
second-order linear differential equation for the leading-order
gluon distribution function directly from the proton structure
function parametrization. The analytical solution of the gluon
distribution for $0<x{\lesssim}x_{P}$ is defined
\begin{eqnarray}
xg(x,Q^2)=-\frac{1}{\omega}\int^{x}\frac{dz}{z}(\frac{z}{x})^{k}\sin{\bigg{(}}\omega
{\ln}(\frac{z}{x})\bigg{)}\mathcal{G}(z,Q^2),
\end{eqnarray}
where $k=-3/2$ and $\omega=\sqrt{7}/2$, the function
$\mathcal{G}(\upsilon,Q^2)$ parametrized in $\upsilon={\ln}(1/x)$
reads
\begin{eqnarray}
\mathcal{G}(\upsilon,Q^2)=\alpha(Q^2)+\beta({Q^2})\upsilon+\gamma(Q^2)\upsilon^{2}.
\end{eqnarray}
The coefficients of the function are quadratic polynomials in $\ln{Q^2}$ [31].\\
%%%%%%%%%%%%%%%%%%%%%%%%%%%%%%%%%%%%%%%%%%%%%%%%%%%%%%%%%%%%%%%%%%%%%%%%
\begin{table}[h]
\caption{ The effective parameters [31] in the domain
$0.11~\mathrm{GeV}^{2}{\leq}Q^{2}{\leq}1200~\mathrm{GeV}^{2}$ and
$10^{-4}{\lesssim}x{\lesssim}0.09$.}
\begin{tabular} {cccc}
\toprule \\  \multicolumn{2}{c}{parameters \quad \quad \quad ~~~~~~~~~~~~~~~~value}    \\ &&&\\ \hline \\ &&&\\
  $a_{0} $  &   \quad  $-5.381\times 10^{-2} \pm  2.17\times10^{-3} $  \\

  $a_{1} $  &   \quad   $2.034\times 10^{-2}~~\pm 1.19\times10^{-3}$  \\

  $a_{2}$   &    \quad  $4.999\times 10^{-4}~~\pm 2.23\times10^{-4}$   \\  &&&\\

 $b_{0}$   &   \quad   $9.955\times 10^{-3}~~\pm 3.09\times10^{-4} $ \\

 $b_{1}$   &   \quad   $3.810\times 10^{-3}~~\pm 1.73\times10^{-4}$  \\

 $b_{2}$    &    \quad  $9.923\times 10^{-4}~~\pm 2.85\times10^{-5} $ \\ &&& \\
\hline
\end{tabular}
\end{table}

%%%%%%%%%%%%%%%%%%%%%%%%%%%%%%%%%%%%%%%%%%%%%%%%%%%

\section{References}
1. K.J.Eskola et al, arXiv[hep-ph]:0110348.\\
2. A. Deshpande, R. Milner, R. Venugopalan and W. Vogelsang, Ann.
Rev. Nucl. Part. Sci. {\bf55}, 165 (2005).\\
3. A. Accardi et al., Eur. Phys. J. A {\bf52}, 268 (2016); R.
Abdul Khalek et al., Nucl. Phys. A {\bf1026}, 122447 (2022).\\
4. J.L. Abelleira Fernandez et al., J. Phys. G: Nucl. Part. Phys.
{\bf39}, 075001 (2012); P. Agostini et al., J. Phys. G: Nucl.
Part. Phys. {\bf48}, 110501 (2021).\\
5. FCC Collaboration (A. Abada et al.), Eur. Phys. J. C {\bf79},
474 (2019); FCC Collaboration (A. Abada et al.), Eur. Phys. J.
Spec. Top. {\bf228}, 755 (2019).\\
6. Anna M.Stasto, Acta Physica Polonica B {\bf16}, 7-{\bf A}23
(2023).\\
7. F.Willeke, Report Number:BNL-221006-2021- FORE,
DOI:10.2172/1765663.\\
8. N.Armesto et al., Phys. Rev. D {\bf105}, 114017 (2023);  N.Armesto, Eur.Phys.J.C {\bf26}, 35 (2002).\\
9. L. V. Gribov, E. M. Levin and M. G. Ryskin, Phys. Rept.
{\bf100}, 1 (1983).\\
10. A. H. Mueller and J.-w. Qiu, Nucl. Phys. B {\bf268}, 427
(1986).\\
11. J.-w. Qiu, Nucl. Phys. B {\bf291}, 746
(1987).\\
12. W. Zhu, Phys.Lett.B {\bf389}, 374 (1996).\\
13. Y. Cai, X.Wang and X.Chen, arXiv [hep-ph]:2401.15651.\\
14. M. Froissart,  Phys. Rev. {\bf123}, 1053 (1961); A. Martin,
Phys. Rev. {\bf129}, 1432 (1963).\\
15. J.Rausch, V.Guzey and M.Klasen, Phys. Rev. D {\bf107}, 054003 (2023).\\
16. P.Duwentaster, V.Guzey, L.Helenius and H.Paukkunen, arXiv:
2312.12993.\\
17. S.Demirci, T.Lappi and S.Schlichting, arXiv[hep-ph]:2312.14585.\\
18. E.R. Cazaroto, F. Carvalho, V.P. Goncalves and F.S. Navarra,
Phys.Lett.B {\bf669}, 331 (2008).\\
19. G. Altarelli and G. Martinelli, Phys. Lett. B {\bf76}, 89 (1978).\\
20. S.Moch,J.A.M.Vermaseren and A.Vogt, Phys.Lett.B {\bf606}, 123
(2005).\\
21. K. J. Eskola, V. J. Kolhinen and C. A. Salgado, Eur. Phys. J.
C {\bf9}, 61 (1999).\\
22. D. de Florian and R. Sassot, Phys. Rev. D {\bf69}, 074028
(2004).\\
23. M. Hirai, S. Kumano and T. H. Nagai, Phys. Rev. C {\bf76},
065207 (2007).\\
24. K. J. Eskola, H. Paukkunen and C. A. Salgado,
arXiv[hep-ph]:0802.0139.\\
25. Wei-tian Deng, Xin-Nian Wang and R. Xu, Phys.Lett.B {\bf701},
133 (2011).\\
26 X.-N. Wang and M. Gyulassy, Phys. Rev. D {\bf44}, 3501 (1991);
Comput. Phys. Commun. {\bf83}, 307 (1994) ; X.-N. Wang, Phys.
Rept. {\bf280}, 287 (1997).\\
27. S. -Y. Li and X. -N. Wang, Phys. Lett. B {\bf527}, 85
(2002).\\
28. W. -T. Deng, X. -N.Wang and R. Xu, Phys.Rev.C {\bf83}, 014915
(2011).\\
29. J. Kwiecinski et al., Phys. Rev. D {\bf42}, 3645 (1990).\\
30. A. Donnachie and P. V. Landshoff Z. Phys. C {\bf61}, 139
(1994); Phys.Lett.B {\bf533}, 277 (2002); Phys.Lett.B {\bf550},
160 (2002); Phys.Lett.B {\bf595}, 393 (2004).\\
31. Martin M. Block, Loyal Durand and Douglas W. McKay, Phys.Rev.D
{\bf77}, 094003 (2008).\\
32. E. L. Berger, M. M. Block and C-I Tan, Phys.Rev.Lett. {\bf98},
242001, (2007).\\
33. K. Kovarik, A. Kusina, T. Jezo,et al. Phys. Rev. D {\bf93},
085037 (2016).\\
34. E. Iancu and R. Venugopalan, arXiv:hep-ph/0303204; A. M.
Stasto, Acta Phys. Polon. B {\bf35}, 3069 (2004); H. Weigert,
Prog. Part. Nucl. Phys. {\bf55}, 461 (2005); J. Jalilian-Marian
and Y. V. Kovchegov, Prog. Part. Nucl. Phys. {\bf56}, 104
(2006).\\
35. M.A.Betemps and M.V.T.Machado, Eur.Phys.J.C {\bf65}, 427
(2010).\\
36. I. I. Balitsky, Nucl. Phys. B {\bf463}, 99 (1996); Phys. Rev.
Lett. {\bf81}, 2024 (1998); Phys. Lett. B {\bf518}, 235 (2001); Y.
V. Kovchegov, Phys. Rev. D {\bf60}, 034008 (1999); Phys. Rev. D
{\bf61}, 074018 (2000).\\

%%%%%%%%%%%%%%%%%%%%%%%%%%%%%%%%%%%%%%%%%%%%%%%%

%%%%%%%%%%%%%%%%%%%%%%%%%%%%%%%%%%%%%%%%%%%%%%%%%

\end{document}